\documentclass[a4paper,11pt]{article}

\usepackage{jcappub} 

\usepackage[T1]{fontenc} 

\usepackage{bm,subfigure,float,siunitx}
\usepackage[figuresright]{rotating}

\newcommand{\be}{\begin{equation}}
\newcommand{\ee}{\end{equation}}

\newcommand{\1}{\left}
\newcommand{\2}{\right}

\newcommand{\dif}{\,\mathrm{d}}

\newcommand{\me}{\mathrm{e}}

\newcommand{\m}{\mu}
\newcommand{\n}{\nu}
\newcommand{\al}{\alpha}
\newcommand{\bet}{\beta}
\newcommand{\gam}{\gamma}

\newcommand{\sig}{\sigma}
\newcommand{\ep}{\epsilon}
\newcommand{\vep}{\varepsilon}

\renewcommand{\th}{\theta}

\title{\boldmath Adiabatic accretion onto black holes in Einstein-Maxwell-scalar theory}

\author[a]{Haiyuan Feng}
\author[a]{Miao Li}
\author[a,1]{Gui-Rong Liang \note{Corresponding author.}}
\author[b]{and Rong-Jia Yang}

\affiliation[a]{Department of Physics, Southern University of Science and Technology, \\Shenzhen 518055, Guangdong, China}
\affiliation[b]{College of Physical Science and Technology, Hebei University, \\Baoding 071002, China}

\emailAdd{406606114@qq.com}
\emailAdd{3498044240@qq.com}
\emailAdd{bluelgr@sina.com}
\emailAdd{yangrongjia@tsinghua.org.cn}

\abstract{We study the adiabatic accretion process of ordinary baryonic gas onto spherically symmetric black holes in Einstein-Maxwell-scalar theory, with two parameters $\alpha$ and $\beta$ in the coupling term. Especially, we demonstrate the range of the transonic points in terms of the charge-to-mass ratio squared and the dimensionless coordinate radius, in two important classes of black holes as examples. Further, we find that the two coupling parameters give modifications to the mass accretion rate at different orders of the sound speed at infinity. We also present their different effects on the temperature ratios of the accreted gas.}

\begin{document}
\maketitle
\flushbottom

\section{Introduction}
The accretion process onto gravitating bodies has demonstrated a broad importance in various astronomical scenarios, such as the formation and evolution of stars, planets, galaxies and galaxy clusters. It also offers the most likely explanation to the high energy outflow from quasars and Active Galactic Nuclei.
Earlier works on accretion done by Hoyle, Lyttleton and Bondi
developed the simplest but standard treatment of the problem by imposing boundary conditions at infinity on the steady state, spherically hydrodynamical flow onto celestial bodies \cite{hoyle_lyttleton_1939,hoyle_lyttleton_1940_1,hoyle_lyttleton_1940_2,1941MNRAS.101..227H,Bondi:1944jm,Bondi:1952ni}, later the corresponding process were fully treated in general relativity \cite{Michel1972}. Among accreting bodies, the black hole solution gives the maximum accretion rate \cite{1983bhwd.book}, thus providing a well-elaborated model and attracting large mounts of theoretical interests.
Further studies were developed in aspects of different black hole backgrounds, various accreted contents, several alternative theories, and in combinations or modifications of them.
In terms of different backgrounds,
accretion onto a Schwarzschild black hole serves as the most classical example and was studied in a broad sense\cite{Font:1998qr,Huerta:2006sc,Mach:2013fsa,Lora-Clavijo:2012oxd,Gangopadhyay:2017tlp,UmarFarooq:2019uqr,Yang:2018zef}; the process by which a gravitating object accretes gas as it moves through a surrounding medium, known as Bondi-Hoyle-Lyttleton accretion, is generalized into several cases \cite{Petrich:1988zz,Jiao:2016uiv,Mach:2020wtm,Yang:2021opo,Mach:2021zqe}; similar procedures were carried out in backgrounds of black holes with rotation or charge \cite{Parev:1995me,Font:1998sc,Medvedev:2002wn,Fragile:2004gp,Babichev:2008dy,Bhadra:2011me,JimenezMadrid:2005rk}, in lower and higher dimensions \cite{Ramirez-Velasquez:2019cgr,Blandford:2003wh,Debnath:2014yna}, and in other types \cite{Jiao:2016iwp, Abbas:2018ygc,Yang:2018cim}. Besides analytical works, more realistic situations were taken into accounts in contexts of astronomical environments \cite{DiMatteo:2002hif,Silich:2008xz,Gil-Merino:2011yhj,Kuo:2014pqa,Kulczycki:2021hcn}.
In terms of different accreted contents, the exact solution was given in dust shell collapsing model \cite{Liu:2009ts}; a different accretion rate dependence was obtained in a string cloud surrounding \cite{Ganguly:2014cqa};
accretions of dark matter and various components of dark energy were also broadly considered \cite{Babichev:2004yx,Babichev:2005py,Babichev:2006xq,Sun:2008si,Gao:2008jv,Martin-Moruno:2009cmc,Sun:2008ev,deLima:2010pt,Sharif:2011zw,Nayak:2011sk,Sharif:2012xn,Kim:2012kh,Amani:2012nt,Debnath:2015yva}.
Generally, the mass of the black hole will decrease during the accretion of dark energy\cite{Babichev:2004yx,Babichev:2005py}, nevertheless, the solution in a Friedmann-Robertson-Walker universe will influence the situation\cite{Gao:2008jv}.
In terms of alternative theories,
the accretion onto black holes were studied in several modified gravity \cite{Sharif:2011ih,Jamil:2011sx,Aslam:2017pxc}, and the correction in quantum gravity was considered \cite{Yang:2015sfa}.

The accretion process for continuous contents onto black holes are well described by
hydrodynamic equations, which are incorporated in the conservation of energy-momentum tensor
\be\label{hydro_tj}
\nabla_\m T^{\m\n}=0,
\ee
where $\nabla_\m$ denotes the covariant derivatives compatible to a given metric $g_{\m\n}$,
and the free index $\n=0$ or $i$ corresponds to the continuity equation of energy or momentum. Its temporal projection gives the continuity equation, while its spatial part is the Euler's equation of motion. For steady state, the energy-momentum tensor is independent of time, so we only have to solve $\nabla_i T^{i\n}=0$. Usually we take the spherical coordinates $(t,r,\th,\phi)$, and for non-rotating radial flow, the equation can be only projected onto two independent spacetime directions,
so we now have two independent formulas in hand.

The mass accretion rate
is defined as the energy flow through the surface of a sphere with a given radius:
\begin{eqnarray}\label{mar}
\label{2}
\dot{M}(r)=\oint T^{1}_{~0}\sqrt{-g}\dif{\theta}\dif{\phi},
\end{eqnarray}
with dot denoting the time derivative. Note that here $M(r)$ doesn't mean the mass of a compact star or a black hole, but the total energy inside the sphere surface with radius $r$; on the consideration of a steady state flow with spherical symmetry, the radial part of
\eqref{hydro_tj} gives
$\partial_r (\sqrt{-g}~ T^1_{~0})=0$,
where a constant of integration is obtained as $C_1=\sqrt{-g}~ T^1_{~0}$,
which further results in a constant mass accretion rate given in \eqref{mar}, so now $M(r)$ is independent of the sphere radius, and we if choose the radius to be near the horizon, we can treat $\dot M$ as the changing rate of black hole mass $M$.

To have an explicit expression for mass accretion rate, we need a specific form of the energy-momentum tensor, generally the ideal fluid case is considered,
\be
T_{\m\n}=(\rho+p)u_\m u_\n+p g_{\m\n},
\ee
where $\rho$ is the energy density, $p$ is the pressure, and $u^\m$ is the $4$-velocity field of the fluid, which obeys the normalization condition $g_{\m\n}u^{\mu}u^{\mu}=-1$.
Specially, for fluid made of particles, we have $\rho=mn+\vep$ with $m$ the particle's rest mass, $n$ the number density, and $\vep$ the internal energy density; in non-relativistic case, we have $p\sim \vep\ll \rho$, this gives the dust energy-momentum tensor $T_{\m\n}=\rho u_\m u_\n$; for Schwarzschild-like metric, $\sqrt{-g}=r^2\sin\th$, we have the good old mass accretion rate as $\dot M=4\pi m n u r^2$, with $u$ denoting the radial velocity.

However, to fully solve the problem, an equation of state of the fluid $p=p(\rho)$ has to be supplied; now we have three independent equations, and three unknowns $p(r), \rho(r), u^\m(r)$, when boundary conditions are specified, an explicit expression for the mass accretion rate can be obtained. Besides, due to volume suppression, the accreted gas is heated during the accretion, and the ratio of the gas temperature near the horizon and that at infinity can be calculated straightforward.

In this paper, we discuss the adiabatic accretion process of non-relativistic baryonic gas onto spherically symmetric black holes in Einstein-Maxwell-scalar (EMS) theory with two parameters $\al$ and $\bet$ in the coupling term, where a non-rotating, steady-state hydrodynamic flow with boundary conditions at infinity is considered. In particular, taking two important classes of black hole solutions as examples, we will present the range of the critical point, i.e., the existence of a stable transonic flow, in terms of the charge to mass ratio squared and the  dimensionless coordinate radius (in the $q^2- r/M$ plane). We will also give different modifications to the mass accretion rate from the two parameters, and show their different influence on the temperature ratios.
The article is organized as follows: In section II, we will derive the fundamental equations for the accretion in a general spherically symmetric spacetime, we will firstly
calculate the mass accretion rate in the corresponding metric,
and then consider the critical points under adiabatic conditions.
In section III, the black hole solutions under the EMS theory will be applied, with the positivity of two critical velocities guaranteed, we present the range in which the critical points exist in two special classes of solutions; then we derive the mass accretion rates for ordinary baryonic matter onto the two classes of black holes, with discussions on the effects from the two coupling parameters; finally we will show the temperature ratio profiles dependent on the two parameters.
We will use geometrical units $c=G=1$ throughout the analysis.

\section{A general analysis on the spherical accretion}

In this section, we demonstrate how the standard procedure works without invoking specific spacetime metric and the equation of state, hence to grasp the ingredients and to give a brief review in the accretion problem. The analytical results will be useful and referred to in the following sections.

\subsection{Mass accretion rate in general spherically-symmetric spacetime}
To apply the procedure in any theory of gravity other than that of Einstein, we consider an arbitrary spacetime which is static and spherically symmetric, with the line element
\be
\label{4}
\dif s^{2}=-f(r)\dif t^{2}+f^{-1}(r)\dif r^{2}+C(r)~r^2 (\dif \theta^2+\sin^2 \theta \dif \phi^2).
\ee
Note that we choose $C(r)$ to be dimensionless,
and assume that the metric is asymptotically flat,  i.e., as $r\rightarrow\infty$, both $f(r)$ and $C(r)\rightarrow 1$,
thus it reduces to a Minkowski metric at spatial infinity.
The volume element is given as $\sqrt{-g}=Cr^2\sin\th$.
Besides, we ignore the back-reaction on the spacetime from the accreted fluid.

We express the radial component of the four-dimensional velocity as $  u^{1}(r)\equiv -v(r) <0$ to express inward flow
, then by normalization we have the contra-variant and covariant $4$-velocity as
$u^\m=(\sqrt{{(1+v^2)}}/f, -v, 0, 0)$ and $u_\m=(-\sqrt{f+v^2}, -Bv, 0, 0)$,
then after integrating over angles, the integration constant $C_1$, induced in the introduction, is given by,
\be
\label{5}
C_{1}
=(\rho+p) u^1 u_0 \sqrt{-g}
=C(\rho+p)vr^2\sqrt{f+v^2},
\ee
then we know the combination $(\rho+p)v$ behaves as $r^{-2}$. One can check that here $C_1$ is also a dimensionless number.

Another independent equation can be obtained by projecting the conservation equation onto the fluid 4-velocity vector, i.e., $u^{\nu}\nabla_{\mu}T^{\mu}_{~\nu} = 0$, by a bit of arranging terms we have
\be
\label{6}
\rho'=-(\rho+p)\frac{\left(Cvr^2\right)'}{Cvr^2}
\ee
where prime denotes the derivative with respect to $r$.
Integrating the above equation gives another integration constant $H$ as
\be
\label{7}
H\equiv Cvr^2~\me^{\left[\int^{\rho}_{\rho_{\infty}}\frac{\dif \rho^{\prime}}{\rho^{\prime}+p(\rho^{\prime})}\right]}
\ee
which is a positive number;
the integration bounds $\rho$ and $\rho_{\infty}$ are the density at $r$ and {at} infinity, respectively. Note that here $H$ has the dimension of length squared. Dividing the two constants in \eqref{5} and \eqref{7} to eliminate $C$, we have the Bernoulli's equation describing the steady accreted fluid along the $r$ coordinate,
\be
\label{8}
C_{2}\equiv\frac{C_{1}}{H}=(\rho+p)\sqrt{f+v^2}~\me^{-\left[\int^{\rho}_{\rho_{\infty}}\frac{\dif \rho^{\prime}}{\rho^{\prime}+p(\rho^{\prime})}\right]},
\ee
which connects the physical conbination concerned at any $r$ with those at infinity serving as our boundary conditions. As $r\rightarrow\infty$, we choose $v\rightarrow 0, \rho\rightarrow \rho_\infty, p\rightarrow p_\infty$, then we immediately have $C_{2}=\rho_{\infty}+p_{\infty}$,
so the mass accretion rate from \eqref{mar} gives
\be
\label{9}
\dot{M}=4\pi C_1=4\pi H C_2
=4\pi H(\rho_{\infty}+p_{\infty})
\ee
It's straightforward to check that both of the left and right hand sides of the equation are dimensionless. Given a specific spacetime metric, and an equation of state, the above formula for the accretion rate of the corresponding gas/fluid can be immediately applied.
We can see the sign of the accretion rate is solely determined by the equation of state; for $\rho_{\infty}+p_{\infty}>0$,
the black hole mass increases as it accretes; while for $\rho_{\infty}+p_{\infty}<0$,
the black hole mass will decrease over time.

\subsection{Conditions for the critical point existence}
In our analysis, we are only concerned with the Bondi accretion, i.e., the fluid accelerates from subsonic to supersonic speeds as it passes through a transonic point or a critical point, which implies a stable flow.
So in this section we specify the conditions in which the critical point exists.
Here we mainly focus on the adiabatic process where the heat taken out by radiation can be neglected.

In the local inertial rest frame of the adiabatic fluid, thermodynamic relations give
\be
\label{13}
0=T\dif s=\dif \left(\frac{\rho}{n}\right)+p\dif \left(\frac{1}{n}\right),
\ee
where we have employed an auxiliary function $n\equiv 1/V$, which is known as the ``concentration" of the fluid, or the particle density in matter made up of particles \cite{Babichev:2005py}. This leads to
\be
\label{14}
\frac{\dif \rho}{\dif n}=\frac{\rho+p}{n},
\ee
which we call it the adiabatic auxiliary relation.
Integrating both sides, we get
\be
\label{17}
\frac{n(r)}{n_{\infty}}=\me^{\int^{\rho}_{\rho_{\infty}}\frac{\dif \rho^{\prime}}{\rho^{\prime}+p(\rho^{\prime})}},
\ee
which indicates the concentration of the fluid changes exponentially with respect to $r$.
This will immediately evaluate the constants $H$ and $C_2$ without knowing the equation of state,
\be\label{HC}\1\{\begin{split}
H&=vCr^2~ \1(\frac n{n_\infty}\2)\\
C_2&=n_\infty\sqrt{f+v^2}\1(\frac{\rho+p}n\2)
\end{split}\2.\ee
and henceforward the mass accretion rate
\be
\label{20}
\dot{M}=4\pi r^{2}Cnv\left(\frac{\rho_{\infty}+p_{\infty}}{n_{\infty}}\right),
\ee
when the values of $n$ and $v$ at arbitrary radius are given, the accretion rate is determined. Later we will use values at the critical point to evaluate this quantity.\\

Recall that $\nabla_\m T^{\m\n}=0$ have only two independent equations in our context. In the previous section, we used $\nabla_\m T^{\m0}$ and $u_\n \nabla_\m T^{\m\n}$ to give eqs.~\eqref{5} and \eqref{6}. To analysis the continuity of the fluid, we want to convert them into two equations in terms of $n(r)$ and $v(r)$ by making use of the auxiliary relation \eqref{14}.

The first independent equation is obtained by inserting the variant of \eqref{14}, $\rho'/(\rho+p)=n'/n$, into eq.~\eqref{6}, which is checked to be the expected particle number conservation,
after expanding we have,
\be\label{par-conser}
\frac{v'}{v}+\frac{n'}{n}=-\frac{(Cr^{2})'}{Cr^{2}}.
\ee
The second equation is nothing but the Euler's equation $\nabla_\m T^{\m1}$, but here we employ the definition of sound speed,
\be
\label{15}
c_{s}^{2}\equiv \frac{\dif p}{\dif \rho}=\frac{n}{\rho+p}\frac{\dif p}{\dif n}=\frac{n}{\rho+p}\frac{p'}{n'},
\ee
with causality constraint $0<c_{s}<1$ assumed.
This leads to
\be
\label{22}
vv'+\frac{n'c_{s}^{2}(f+v^2)}{n}=-\frac{1}{2}f'.
\ee
Joining eqs.~\eqref{par-conser} and \eqref{22}, we can solve for $v'$ and $n'$ as
\be
\1\{\begin{split}
\label{23}
v'&=\phantom{-}\frac{D_{1}}{D}\\
n'&=-\frac{D_{2}}{D},
\end{split}\2.
\ee
with the numerators and the denominators given as
\be
\1\{\begin{split}
\label{24}
D_{1}&=\frac{1}{n}\left[\frac{c_{s}^{2}(f+v^2)(Cr^{2})'}{Cr^{2}}-\frac{1}{2}f' \right]\\
D_{2}&=\frac{1}{v}\left[\frac{v^2(Cr^{2})'}{Cr^{2}}-\frac{1}{2}f'\right] \\
D&=\frac{v^2-c_{s}^{2}(f+v^2)}{nv}.
\end{split}\2.
\ee
Continuity constraint of the fluid demands $v'$ and $n'$ are nonsingular everywhere in spacetime, which implies
that if the denominator in eq.~\eqref{23} is zero at some point, the numerator must also be zero. 
This point is called the critical point, with the corresponding radius determined by $D=D_{1}=D_{2}=0$. In the following, we denote the corresponding values at the critical point with the subscript ``$c$". The derivation results in
\be
\1\{\begin{split}
\label{25}
c_{sc}^{2}&=\frac{f_c'}{f_c'+2 f_c/\Omega_c}\\
v^{2}_{c}&=\frac{f_c'}2~\Omega_c,
\end{split}\2.
\ee
with the geometric factor defined as
\be
\Omega_c\equiv\1[\frac{C r^{2}}{(C r^{2})'}\2]_c,
\ee
and the relation between the two velocities is given by
\be
v^{2}_{c}=\frac{c_{sc}^{2}f_c}{1-c_{sc}^2}\\.
\ee
It is straightforward to check that
in Schwarzschild spacetime, where $f(r)=1-{2M}/{r}$, and $C(r)=1$, the above formula gives $v^{2}_{c}
={c_{sc}^{2}}/{(1+3c_{sc}^{2})}=M/(2r_c)$, which is consistent with the result in \cite{1983bhwd.book}. The conditions for critical point to exist is that both $v^{2}_{c}$ and $c_{sc}^{2}$ are positive. This would constrain the position of the critical radius, and possibly other parameters in an arbitrary gravitational theory, e.g., the charge mass ration in EMS theory, which will be discussed in the next section.

Further, with the help of an explicit equation of state, the value of $n_c$ can be calculated, then we insert values of $v_c$ and $n_c$ and a boundary condition into eq.~\eqref{20}, we get the expression of the mass accretion rate.

\section{Accretion process in Einstein-Maxwell-scalar theory}
Einstein-Maxwell-scalar theory, which arises as a low-energy effective theory in string theory,
extends the Einstein-Maxwell theory by including the kinetic term associated to a background scalar field and the coupling to the Maxwell term \cite{Yu:2020rqi,Qiu:2021qrt}, with its special case known as the Einstein-Maxwell-dilaton (EMD) theory \cite{GoulartSantos:2017dun,Heydari-Fard:2020ugv}.
The action of EMS theory has the following form:
\be
\label{26}
S[g_{\mu\nu},A_{\mu},\phi]=\frac{1}{16\pi}\int d^{4}x\sqrt{-g}\1[R-2g^{\mu\nu}\partial_{\mu}\phi\partial_{\nu}\phi-K(\phi)F^{2}\2],
\ee
where $R$ is the Ricci scalar, $\phi$ is the scalar field, $F^{2}\equiv F^{\mu\nu}F_{\mu\nu}$ with $F_{\mu\nu}=\partial_{\mu}A_{\nu}-\partial_{\nu}A_{\mu}$ the electromagnetic field tensor, and the coupling function $K(\phi)$ is given by \cite{Yu:2020rqi},
\be
K(\phi)=\frac{(\al^{2}+1)e^{\frac{-2\phi}{\al}}}{(\al^{2}+1+\beta)e^{\frac{-2\phi(\al^{2}+1)}{\al}}+\beta \al^{2}},
\ee
with $\al$ and $\bet$ two characteristic coupling parameters in the theory; it is seen that as $\beta\rightarrow\infty$, the coupling from the electromagnetic field will be ignored, and another extremal case is that when $\bet=0$, it reduces to the EMD coupling with $K(\phi)=\me^{2\al\phi}$.

The variation of the EMS action \eqref{26} yields three equations of motion:
\be
\1\{\begin{split}
\label{27}
0&=\nabla_{\mu}\left[K(\phi)F^{\mu\nu}\right]\\
\Box\phi&=\frac{1}{4}\frac{\partial K(\phi)}{\partial \phi}F^{2}\\
R_{\mu\nu}&=2\partial_{\mu}\phi\partial_{\nu}\phi+2K(\phi)\left(F_{\mu\sig}F^{~\sig}_{\nu}-\frac{1}{4}g_{\mu\nu}F^{2}\right).
\end{split}\2.
\ee
The general spherically symmetric solution
has the following form \cite{Qiu:2021qrt}:

\be\label{28}
\1\{\begin{split}
f(r)&=\left(1-\frac{b_{1}}{r}\right)\left(1-\frac{b_{2}}{r}\right)^{\frac{1-\al^{2}}{1+\al^{2}}}+\frac{\beta Q^{2}}{r^{2}C(r)}\\
C(r)&=\left(1-\frac{b_{2}}{r}\right)^{\frac{2\al^{2}}{1+\al^{2}}},
\end{split}\2.
\ee
where $b_{1}$ and $b_{2}$ are only functions of $\al$, they are given as
\be
\1\{\begin{split}
\label{29}
b_{1}&=(1+\sqrt{1-q^{2}(1-\al^{2})})~M\\
b_{2}&=\frac{1+\al^{2}}{1-\al^{2}}\left[1-\sqrt{1-q^{2}(1-\al^{2})}\right]~M.
\end{split}\2.
\ee
with $q\equiv{Q}/{M}$ the charge-to-mass ratio, and $M$ the mass of the black hole. The location of horizons are given by $f(r_{\pm})=0$; then if $\bet=0$, the solutions are known as the GMGHS solutions \cite{Garfinkle:1990qj}, $r_{\pm}$ are just $b_1$ and $b_2$; and further if $\al=0$, they are horizons of the Reissner-Nordstrom black hole, and when $q=0$ they
reduce to the Schwarszchild case. It is noted that in the ``sGMGHS-like" case (we name it, ``s" for ``special", by fixing $\al=1$; and ``-like" for a whole class of allowable $\bet$), there are two horizons only in the $0<q<\sqrt{2},~ 0<\beta<\frac{(2-q^2)^2}{4q^2}$ region \cite{Yu:2020rqi}.

For physical clarity and analytical convenience, in the following discussions we mainly focus on two classes of solutions: (1) the GMGHS solutions, with $\bet=0$ but a free choice of $\al$; (2) the sGMGHS-like solutions, with $\al=1$ but different values of $\bet$.

\subsection{Constraints on the critical radius in GMGHS and sGMGHS-like black holes}

As noted before, the positivity of $v^{2}_{c}$ and $c_{sc}^{2}$ would constrain the range of critical radius $r_c$ and the charge mass ratio $q$ in EMS theory. In this section, we will determine this range, in GMGHS cases ($\bet=0$) and in sGMGHS-like cases ($\al=1$), respectively.
We expect that as $q$ varies, $r_c$ of a stability flow will also change. According to eq.~\eqref{9}, there are two possibilities for $q$'s evolution, depending on the equation of state of the accreted flow.
When studying non-relativistic baryonic gas, the black hole mass will increase, making $q$ to decrease;
when considering certain types of dark energy (e.g. phantom dark energy) satisfying
$p+\rho<0$, the black hole mass will decrease, making $q$ to increase. In the later case, there would be a chance for a large enough $q$
breaking the weak cosmic censorship conjecture and leading to
a naked singularity. In the GMGHS cases ($\bet=0$), the criterial condition for the weak cosmic censorship conjecture to be protected is $q^{2}<1+\al^{2}$;
we will obtain $\al=0,1$ and $\sqrt{3}$ constraints respectively. In the sGMGHS-like cases ($\al=1$), the criteria is $q^{2}<2$ and $\bet<\frac{(2-q^2)^2}{4q^2}$, which can be solved as $q^2<2[(\bet+1)-\sqrt{(\bet+1)^2-1}]<2$, indicating that a non-zero $\bet$ would narrower the range of $q^2$ which preserves the weak cosmic censorship, with a smaller upper bound as the value of $\bet$ becomes larger; in the following we will show $\bet=0.1, 0.5, 1$ results as examples.

The explicit expressions for the critical velocity and critical sound velocity of the ideal fluid  are given by substituting the metric \eqref{28} into \eqref{25},
\be
\1\{\begin{split}
\label{vc_tot}
v^2_{c}&=\frac{2 \beta  Q^2 \left[b_2-\left(\alpha ^2+1\right) r_c\right]+\left(r_c-b_2\right) \left[\left(\alpha^2+1\right) b_2r_c+\left(\alpha ^2+1\right) b_1 r_c-2 b_1 b_2\right]}{4 r_c^2 \left[\left(\alpha ^2+1\right) r_c-b_2\right] \left(1-\frac{b_2}{r_c}\right){}^{\frac{2 \alpha ^2}{\alpha ^2+1}}}\\
c^2_{sc}&=\frac{2 \beta  Q^2 \left[b_2-\left(\alpha ^2+1\right) r_c\right]+\left(r_c-b_2\right) \left[\left(\alpha^2+1\right) b_2r_c+\left(\alpha ^2+1\right) b_1 r_c-2 b_1 b_2\right]}{2 \beta  Q^2 \left[\left(\alpha ^2+1\right) r_c-b_2\right]+\left(r_c-b_2\right) \left[-3 \left(\alpha ^2+1\right) b_1 r_c-\left(\alpha ^2+3\right) b_2 r_c+2 b_1 b_2+4 \left(\alpha ^2+1\right) r_c^2\right]}.
\end{split}\2.
\ee
It is checked that they return to the expressions in RN case when $\al=0$ and $\bet=0$, consistent with that given in \cite{Jamil:2008bc}.

\begin{figure}[tbp]
\centering
\includegraphics[scale=1]{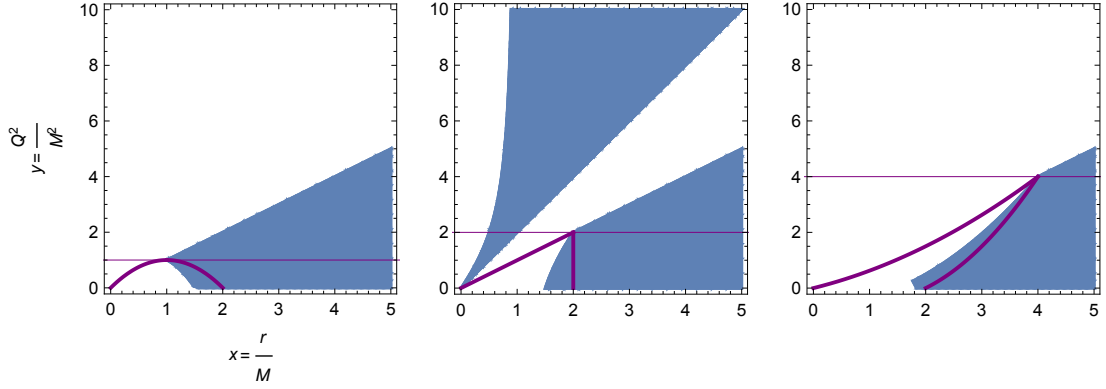}
\caption{\label{EMDcrit}The validity region for a stable flow existence in GMGHS cases ($\bet=0$), with three sub-cases $\al=0$ (left) $\al=1$ (middle) and $\al=\sqrt{3}$ (right) shown. The regions are determined in the blue area, with the horizontal axis $x$ representing the ratio of the radius to mass $r/M$, and the vertical axis $y$ representing the square of the charge-to-mass ratio $q^2$. The inner and outer event horizons are denoted by thick purple lines, and the extremal black holes denoted by horizontal thin purple lines are given as $y=1+\al^2=1,2,4$ for $\al=0,1,\sqrt{3}$ sub-cases.}
\end{figure}

\begin{figure}[tbp]
\centering
\includegraphics[scale=1]{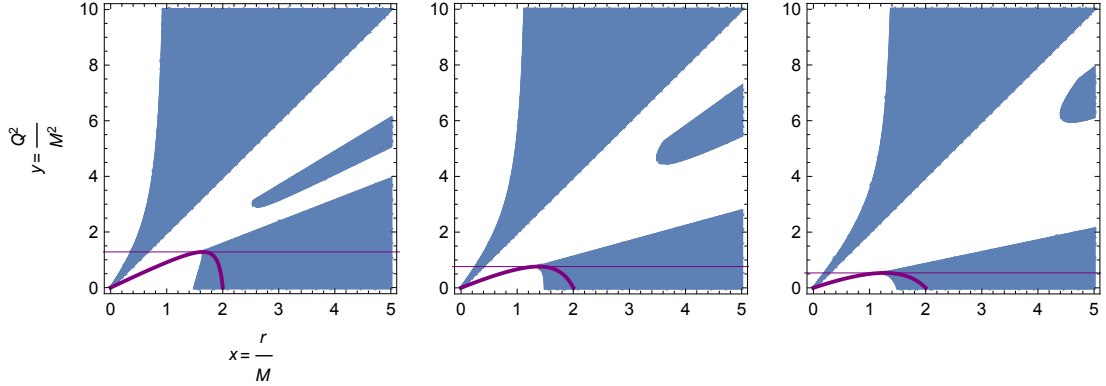}
\caption{\label{EMScrit}The validity region for a stable flow existence in sGMGHS-like cases ($\al=1$), with three sub-cases $\bet=0.1$ (left) $\bet=0.5$ (middle) and $\bet=1$ (right) shown. The regions are determined in the blue area, with the horizontal axis $x$ representing the ratio of the radius to mass $r/M$, and the vertical axis $y$ representing the square of the charge-to-mass ratio $q^2$. The inner and outer event horizons are denoted by thick purple lines, and the horizontal thin purple lines $y=2[(\bet+1)-\sqrt{(\bet+1)^2-1}]=1.28,0.76,0.54$ for $\bet=0.1,0.5,1$ represent the upper limits of the charge-to-mass ratio which preserves the weak cosmic censorship conjecture.}
\end{figure}

For convenience, we introduce two dimensionless variables: the coordinate distance in unit of $M$ as $x\equiv{r}/{M}$, and the squared charge-to-mass ratio $y\equiv q^2$.
Then the event horizons in GMGHS case are given as
$x_{\pm}=b_{1,2}/M$,
which are two sectors of a parabola, meeting at the extremal case $y=1+\alpha^{2}$.
With the positivity
$v^{2}_{c}\geq0$, and $c_{sc}^{2}\geq0$ guaranteed,
the ranges of parameters $r_c/M$ and $q^2$ in GMGHS cases are shown in the blue area in figure~\ref{EMDcrit}, and the ranges in sGMGHS-like cases are shown in figure~\ref{EMScrit}.


For a given initial state of the black hole, say $0<q<q_{\text{max}}$ (imagining a horizontal line in region $0<y<1+\al^2$ in GMGHS case or $0<y<2[(\bet+1)-\sqrt{(\bet+1)^2-1}]$ in sGMGHS-like case), the accreting of baryonic matter would make $q$ decrease; while accreting of dark energy, if satisfying $p+\rho<0$, would make $q$ to increase, with the chance of breaking the weak cosmic censorship conjecture. Note that here the validity of a stable flow does not give the criteria whether the weak cosmic censorship conjecture is preserved or broken, it only tells that in this region we have a transonic flow solution.

\subsection{Mass accretion rate and temperature ratio in the accretion of baryonic gas}

With a stable flow determined, we can discuss the accretion in terms of various accreted contents, with different equation of state. Here we study the non-relativistic baryonic gas, with polytropic equation of state given as
\be
\label{36}
p=kn^{\gamma},
\ee
where $\gamma$ is adiabatic index
and $k$ is a constant. Substitute the above formula into \eqref{14}  and integrate, one can get
\be
\rho=mn+\frac{p}{\gamma-1},
\ee
with $m$ the integration constant interpreted as the rest mass of a particle. This is consistent with that given in the introduction, with identifying the internal energy density as $\ep={p}/{(\gamma-1)}$. Differentiating both sides with respect to $p$ and invoking the definition of sound speed, we have
\be
\frac1{c_{sc}^2}=\frac{\dif \rho}{\dif p}=m \frac{\dif n}{\dif p}+\frac{1}{\gamma-1},
\ee
after extracting out $\dif p/\dif n$, we can express the combination $(\rho+p)/n$ fully in terms of $c_s^2$,
\be
\frac{\rho+p}n=\frac{\dif \rho}{\dif p}\frac{\dif p}{\dif n}=\frac1{c_{sc}^2} \frac{m}{\frac1{c_s^2}-\frac1{\gam-1}}=\frac m {1-\frac{c_s^2}{\gam-1}},
\ee
then the equality for $C_2$ in \eqref{HC} gives the Bernoulli equation 
as
\be
\label{38}
\left(1+\frac{c_{s}^{2}}{\gamma-1-c_{s}^{2}}\right)^{2}\left[f(r)+v^{2}\right]
=\left(1+\frac{c_{s\infty}^{2}}{\gamma-1-c_{s\infty}^{2}}\right)^{2},
\ee
we write out explicitly the expression in EMS theory,
\be\label{bnl_ems}
\left(1+\frac{c_{s}^{2}}{\gamma-1-c_{s}^{2}}\right)^{2}\left[\left(1-\frac{b_{1}}{r}\right)\left(1-\frac{b_{2}}{r}\right)^{\frac{1-\al^{2}}{1+\al^{2}}}+\frac{\beta Q^2}{r^{2}C(r)}+v^{2}\right]=\left(1+\frac{c_{s\infty}^{2}}{\gamma-1-c_{s\infty}^{2}}\right)^{2}.
\ee
This equation, obtained from the equation of state, together with those two in \eqref{vc_tot}  originated from the energy-momentum conservation, fully determine the three unknowns $v_c$, $c_{sc}$ and $r_c$ in both cases. However the solutions are tedious, here we impose some approximations to simply the expressions.

We investigate the case in which the critical point is far away from the horizon, i.e.,
$r_c\gg r_+$, {which corresponds to a non-relativistic flow almost uninfluenced by the nonlinear gravity;} thus we have $v_c\ll 1$, $c_{sc}\ll 1$, so that in EMS theory \eqref{vc_tot} and \eqref{bnl_ems} can be approximated to
\be
\1\{\begin{split}
&r_{c}\approx\frac{M}{2c^{2}_{sc}}+\frac{3M}{2}+\frac{Q^{2}}{M}\1[\frac{\al^{2}}{1+\sqrt{1-q^{2}(1-\al^{2})}}-\beta-1\2]\\
&v^{2}_{c}\approx c_{sc}^{2}\\
&c^{2}_{sc}\approx\frac{2c_{s\infty}^{2}}{5-3\gamma},
\end{split}\2.
\ee
we see that $\al$ and $\bet$ give a correction term to the Schwartzchild case respectively,  
while the last two equations hold the same form. Now we expressed the critical values $r_{c}$, $v_{c}$ and $c_{sc}$ in terms of the boundary sound speed $c_{s\infty}$.

Further, with the help of the equation of state, we can also have the particle number density $n$ in terms of the sound speed $c_{s}$, as
\be
\gamma k n^{\gamma-1}=\frac{mc_{s}^{2}}{1-\frac{c_{s}^{2}}{\gamma-1}},
\ee
and the critical value of the number density is expressed in terms of the boundary value as
\be
\label{41}
n_{c}\approx n_{\infty}\left(\frac{c_{sc}}{c_{s\infty}}\right)^{\frac{2}{\gamma-1}}=n_{\infty}\left(\frac{2}{5-3\gamma}\right)^{\frac{1}{\gamma-1}},
\ee
this equation also holds the same form, with no corrections from $\al$ or $\bet$.

By using all the above, we insert critical values into \eqref{20} to give the black hole mass accretion rate in EMS theory as
\be
\label{40}
\begin{split}
\dot{M}
&=4\pi r^{2}_{c}\left(1-\frac{b_{2}}{r_{c}}\right)^{\frac{2\al^{2}}{1+\al^{2}}}mn_{c}v_{c}\\
&=4\pi M^{2}mn_{\infty}c_{s\infty}^{-3}\left(\frac{1}{2}\right)^{\frac{\gamma+1}{2(\gamma-1)}}\left(\frac{5-3\gamma}{4}\right)^{\frac{3\gamma-5}{2(\gamma-1)}}g(q,\al,\beta,\gamma),
\end{split}
\ee
where the dimensionless correction factor $g(q,\al, \bet, \gamma)$ is calculated as
\be
\begin{split}
g(q,\al,\beta,\gamma)&\equiv
\1(\frac{4c_{sc}^4}{M^2}\2) r_c^2~ C(r_c) \\
&= \1\{1+c_{sc}^2\1[3-2q^2(1+\bet)+\frac{2q^2\al^{2}}{1+\sqrt{1-q^{2}(1-\al^{2})}}\2]\2\}^2\left(1-\frac{b_{2}}{r_{c}}\right)^{\frac{2\al^2}{1+\al^2}} \\
&\approx1+c^{2}_{sc}\1[{6-4q^{2}(1+\beta)}\2]+O(c^4_{sc})\\
&=1+c^{2}_{s\infty}\frac{12-8q^{2}(1+\beta)}{5-3\gamma}+O(c^4_{s\infty}).
\end{split}
\ee
In the above equation, we expanded the correction factor $g(q,\al,\beta,\gamma)$ in terms of different orders of $c_{sc}$ or finally of $c_{s\infty}$. We see that at zeroth order, it agrees with the results in Schwartzchild case \cite{1983bhwd.book}; at the level of $c_{s\infty}^2$, the influence of the charge-to-mass ratio $q$ and the second parameter $\bet$ shows up, when $\bet=0$ it is consistent with the RN results \cite{FreitasPacheco:2011yme}; the modification due to $\al$ only occurs at the order of $c_{s\infty}^4$, thus is small enough and can be neglected. This indicates that $\bet$ does play a more important role than $\al$ in EMS accretion, while the EMD ($\bet=0$) accretion results
are the same as that of RN case, up to the order of $c_{s\infty}^2$. We also note that this correction is purely geometrical, which is clearly seen that $r^2 C(r)$ originates from the line element of the sphere.

\begin{figure}[tbp]
\centering
\begin{minipage}{0.5\textwidth}
\centering
\includegraphics[scale=0.8,angle=0]{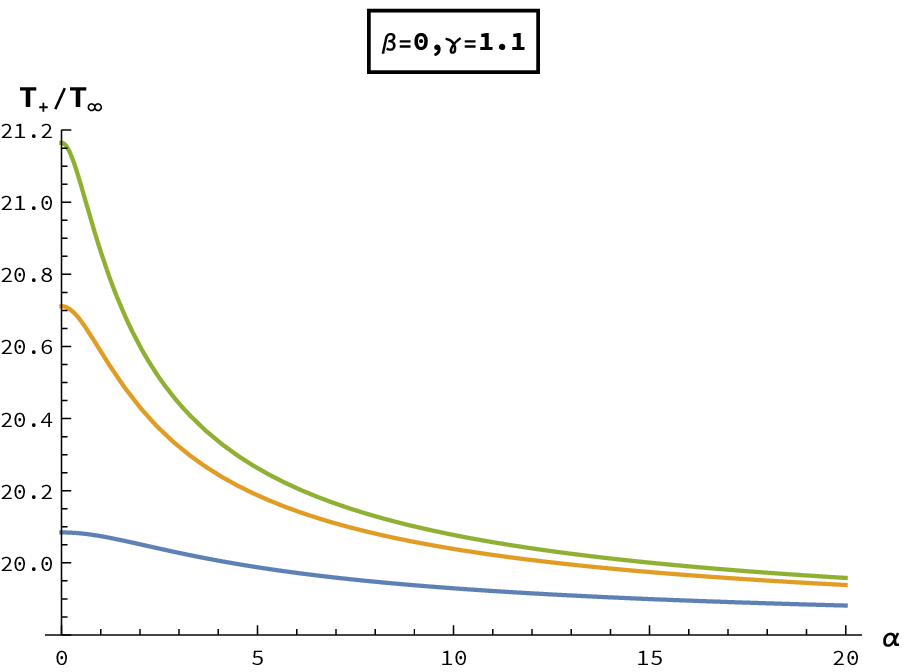}
\end{minipage}%
\begin{minipage}{0.5\textwidth}
\centering
\includegraphics[scale=0.8,angle=0]{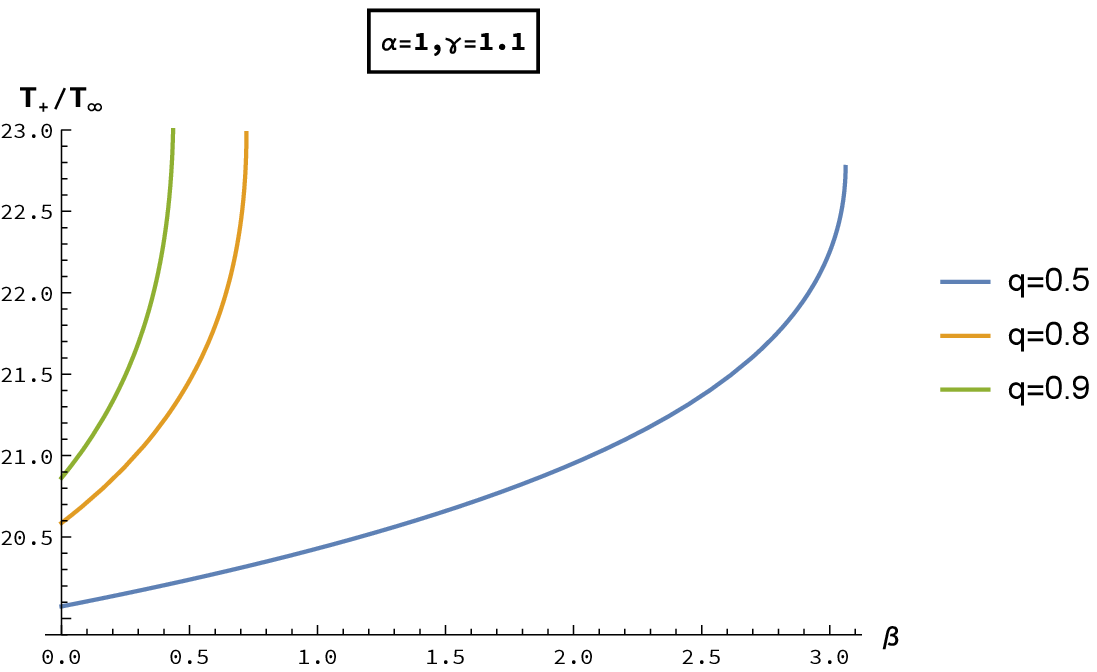}
\end{minipage}
\caption{\label{tem_ratio}The ratios between the temperatures of the accreted gas near the horizon $T_+\equiv T(r_+)$ and that at infinity $T_\infty$, as functions of the coupling parameter $\al$ in GMGHS case (left) and of $\beta$ in sGMGHS-like case (right). We fix the adiabatic index as $\gamma=1.1$, and demonstrate the relations with $q=0.5, 0.8, 0.9$, denoted by blue, orange and green lines. The sound speed at infinity $c_{s\infty}$ is chosen to be $1/{30000}$, taken from \cite{Sharif:2016pqy}.}
\end{figure}

Assuming that the baryonic gas satisfies Maxwell-Boltzmann distribution $p=nk_{B}T$, then by using the equation of state \eqref{36} and the relation \eqref{41}, we get the ratio of the adiabatic temperature near the horizon and to that in the infinity as
\be
\label{45}
\frac{T(r_+)}{T_{\infty}}=\left[\frac{n(r_+)}{n_{\infty}}\right]^{\gamma-1}=\frac2{5-3\gam}\left[\frac{n(r_+)}{n_{c}}\right]^{\gamma-1},
\ee
where the number density ratio can be obtained by invoking the expression of the mass accretion rate \eqref{40}. We approximate the flow velocity near the outer horizon as $v(r_+)\simeq 1$, then we have the temperature ratio in the EMS theory as
\be\begin{split}
\label{Temd}
\frac{T(r_{+})}{T_{\infty}}&=\frac2{5-3\gam}\1[\frac{r_c^2 ~C(r_c) ~c_{sc}}{r_+^2 C(r_+)}\2]^{\gam-1}=\1(\frac2{5-3\gam}\2)^{\frac{\gam+1}2}\1[\frac{r_c^2 ~C(r_c) ~c_{s\infty}}{r_+^2 C(r_+)}\2]^{\gam-1}\\
&=\left(\frac{1}{2}\right)^{\frac{\gamma+1}{2}}\left(\frac{5-3\gamma}{4}\right)^{\frac{3\gamma-5}{2}}\1[\frac{ M^2 g(q,\al,\bet,\gamma)}{ ~c_{s\infty}^3r_+^2 \left(1-\frac{b_2}{r_{+}}\right)^{\frac{2\al^2}{1+\al^2}}}\2]^{\gam-1}.
\end{split}\ee
We see that since $g(q,\al,\bet,\gamma)\simeq 1$, this ratio is solely determined by the black hole surface area proportional to $r^2_+ C(r_+)$. We plot the $\al$ and $\bet$-dependent temperature ratios in figure~\ref{tem_ratio}.

It is seen that there's a major difference in the varying trend of the temperature ratio with respect to $\al$ and $\bet$. As $\al$ goes up, the ratio decreases, while as $\bet$ goes up, the ratio increases. 
Note that the cosmic censorship conjecture $q^2<1+\al^2$ is always satisfied in the left panel of the figure, so there's no divergent point; however, the same constraint gives $\bet<\frac{(2-q^2)^2}{4q^2}$, thus making the ratio blow up at large enough $\bet$, for $q=0.5$, $0.8$, and $0.9$, the blow up points locate at $\bet\simeq 3.063$, $0.723$, and $0.437$, respectively.

Also note that the temperature ratios is sensitive to the adiabatic index $\gam$, according to our analytical formulas, they increases by roughly $100$ times as $\gam$ goes up by $0.2$; but the curve profiles keep the same as that in the case of small $\gam$, so we didn't plot the $\gam$-dependence of temperature ratios in the figures.

\section{Conclusion and discussion}
In this work, we studied the adiabatic accretion process of ordinary polytropic baryonic gas onto spherically symmetric black holes in Einstein-Maxwell-scalar theory, by invoking a solution with two parameters $\al$ and $\bet$ in the coupling term. As usual, we reviewed and applied the standard procedure of accretion problem, by considering a non-rotating, steady-state hydrodynamic flow with boundary conditions at infinity. We determine the ranges of the transonic points, and calculated the mass accretion rates and the temperature ratios of the accreted gas. Two novel results are obtained in our research.

Firstly, due to the positivity of the critical flow velocity and sound speed, the position of the critical radius, or the transonic point, is not arbitrary. The existence of a transonic point is necessary for a physically stable transonic flow, thus it's a premise for the later analysis. Since in the black hole solution under EMS theory there are two variables that can influence the transonic position --- the charge-to-mass ratio and the coordinate radius, we determined the range of the points in terms of them (in the $q^2- r/M$ plane). As examples, the results for two classes of black hole solutions are presented in the figures, where we also indicated the horizons and the extremal line. It is shown that there may be several disconnected patches which provided the range for the transonic point existence, further with a possibility that some portions of these patches exceed the allowance of the cosmic censorship conjecture. We note that these analytical results cannot be criteria to judge the cosmic censorship conjecture, which leaves space for the study of accretion of certain types of dark energy in the future.

Secondly, we calculated the mass accretion rates and the gas temperature ratios in the theory. We find that the two coupling parameters give modifications to the mass accretion rate at different orders of the sound speed at infinity, with $\bet$ contributing at $c_{s\infty}^2$, while $\al$ firstly appears only at $c_{s\infty}^4$. For the gas temperature ratio, we obtain the analytical expressions and demonstrate the result in GMGHS and sGMGHS-like case respectively, in consistency with the preceding analysis on the transonic flow. It is shown that the two parameters modify the Schwartzchild result in different ways, with the increase of $\al$, the ratio decays and tends to a stable value, while with a larger value of $\bet$, the ratio enhances significantly and finally diverges at the extremal point. The above different parameter dependence can be partly traced back to the expression of the critical point, and it also originates from the different status of parameters in the metric.

The present study can be readily generalized to several direction. The accretion of certain types of dark energy would make the mass accretion rate negative, which gives a chance to challenge the cosmic censorship conjecture; how the initial states of a black hole and parameters in the theory, including possibly non-zero cosmological constants, influence the process would be an interesting topic to investigate. It is also straightforward to generalize the analysis to cases of rotating or moving black holes immersed in rotating gas in EMS theory. However, from the study of effects of parameters, what mostly fascinates us is if we can extract some universal ingredients from a general metric in different theories and obtain more insights without detailed computations. We will report our studies along this routine afterwards. Besides, we will try to study gravitational waves produced and propagating in the accretion process in different theories.

\acknowledgments
We thank Prof. Hongbao Zhang for useful discussions and suggestions. The work is in part supported by National Natural Science Foundation of China under Grants No. 12175099 (M.Li) and No. 12147163 (G-R. Liang), and Hebei Provincial Natural Science Foundation of China under Grant No. A2021201034 (R-J. Yang).

\bibliographystyle{JHEP}
\bibliography{accrefs}
\end{document}